\documentstyle[epsf,psfig]{elsart}
\begin{document}
\hyphenation{sub-milli-me-tre as-tro-no-mi-cal or-der}
\newcommand{\eg}{{\sl e.g. }}
\newcommand{\ie}{{\it i.e. }}
\newcommand{\micron}{\hbox{$\,\mu {\rm m}$}}
\newcommand{\microK}{\hbox{$\,\mu {\rm K}$}}
\newcommand{\milli}{\hbox{$\, {\rm mm}$}}
\newcommand{\metre}{\hbox{$\, {\rm m}$}}
\newcommand{\zu}{\rm\,}     
\newcommand{\nanom}{\hbox{$\,{\rm  nm}$}}
\newcommand{\degree}{\hbox{$\,^{\rm  \circ}$}}
\newcommand{\Wms}{\hbox{$\,{\rm  W\, m^{-2}\, sr^{-1}}$}}
\newcommand{\nWms}{\hbox{$\,{\rm  n\!W\, m^{-2}\, sr^{-1}}$}}
\newcommand{\Hcm}{\hbox{$\,{\rm  H\, cm^{-2}}$}}
\newcommand{\Kkms}{\hbox{$\,{\rm  K\, km\,s^{-1}}$}}
\newcommand{\Htwo}{$\rm  H_2$}
\newcommand{\cf}{{\it c.f. }}
\newcommand{\teqline}{\medskip\hrulefill\medskip}
\newcommand{\tabline}{\noalign{
     \medskip}\noalign{\hrule}\noalign{\medskip}}
\newcommand{\om}{\omit&}
\begin{frontmatter}
\title{Observations of the Sunyaev--Zel'dovich effect \\ at high angular
resolution towards \\the galaxy clusters A665, A2163 and CL0016+16}
\author[LAOG,IAS]{F.-X. D\'esert},
\author[CRTBT]{A. Benoit},
\author[CESR]{S. Gaertner},
\author[IAS]{J.--P. Bernard},
\author[IAS]{N. Coron},
\author[IAS,EFI]{J. Delabrouille},
\author[IAS]{P. de Marcillac},
\author[CESR]{M. Giard} 
\author[IAS]{J.--M. Lamarre},
\author[IRAM]{B. Lefloch},
\author[IAS]{J.--L. Puget},
and
\author[CRTBT]{A. Sirbi}
\address[LAOG]{Laboratoire d'Astrophysique de l'Observatoire de Grenoble, 414 rue de la Piscine, BP 53,
F--38041 Grenoble Cedex 9}
\address[IAS]{Institut d'Astrophysique Spatiale, B\^at. 121,
Universit\'e Paris XI, F--91405 Orsay Cedex France}
\address[CRTBT]{ Centre de Recherche sur les Tr\`es Basses
Temp\'eratures, 25 Avenue des Martyrs BP166, F--38042 Grenoble Cedex 9
France}
\address[IRAM]{IRAM, avd Divina Pastora 7 Nucleo Central 18012 Granada Spain}
\address[CESR]{Centre d'\'Etudes Spatiales sur les Rayonnements, 9 Avenue du
Colonel Roche, BP 4346, F--31029 Toulouse Cedex France}
\address[EFI]{  Enrico Fermi Institute, University of Chicago, 
5460 South Ellis Avenue, Chicago, IL 60637, USA}
\begin{abstract}
  We report on the first observation of the Sunyaev--Zel'dovich (SZ)
  effect, a distortion of the Cosmic Microwave Background radiation
  (CMB) by hot electrons in clusters of galaxies, with the Diabolo
  experiment at the IRAM 30\metre{} telescope. Diabolo is a
  dual--channel 0.1K bolometer photometer dedicated to the observation
  of CMB anisotropies at 2.1 and 1.2\milli{}.  A significant
  brightness decrement in the 2.1\milli{} channel is detected in the
  direction of three clusters (Abell 665, Abell 2163 and CL0016+16).
  With a 30 arcsecond beam and 3 arcminute beamthrow, this is the
  highest angular resolution observation to date of the SZ effect.
  Interleaving integrations on targets and on nearby blank fields have
  been performed in order to check and correct for systematic effects.
  Gas masses can be directly inferred from these observations.
\end{abstract} 
\end{frontmatter} 

\section{Introduction}
After the discovery of the Cosmic Microwave Background (CMB) radiation
by Penzias \& Wilson~\cite{Penz65}, and the observation of hot ionised
gas in clusters of galaxies through its X--ray emission~\cite{Lea73},
Sunyaev \& Zel'dovich~\cite{Suny70} soon realised that the scattering
of the CMB photons by the hot electrons of the intracluster medium
(ICM) should generate a distinctive spectral distortion of the CMB
blackbody spectrum in the (sub)millimetre and radio domain.  Several
millimetre and radio detections towards a dozen of clusters have
recently been obtained using various techniques
\cite{Birk91a,Birk91b,Carl96,Jone93,Wilb94,Herb95}. These results,
which are compatible with the expected brightness decrement,
constitute a direct evidence for the SZ effect and have profound
cosmological importance:
\begin{itemize}
\item They are a strong confirmation of the cosmological origin of the CMB
radiation.
\item The mass of the ionised gas in
clusters of galaxies can be obtained from SZ 
measurements, even for unresolved clusters~\cite{Delu95}. If hydrostatic 
equilibrium is assumed, the total mass can also be deduced from the SZ 
profile, and compared
with cluster mass estimates by other methods (gravitational lensing, velocity
fields) for consistency. This, together with cluster number counts, yields
an estimate of $\Omega$ at cluster scales.
\item The detection via the SZ effect of very distant clusters ($z \simeq 1$
and above) would put severe constraints on $\Omega$, as only in a low-density
Universe could structures form so early (\eg~\cite{Barb96}).
\item The angular diameter distance to a cluster can be estimated from the
CMB intensity change due to the SZ effect combined with the observed X-ray
surface brightness. For low redshift clusters, the combination of SZ and X-ray
data thus allows estimating the Hubble constant 
$H_0$~\cite{Birk79,Cava79,Silk78}. For high redshift clusters, because of the
additional dependence of the angular diameter distance on the deceleration
parameter $q_0$~\cite{Silk78}, it is also possible, in principle, to constrain
$\Omega$ and $\Lambda$ (see~\cite{Koba96} for an application of the
method to available SZ measurements).
\item The measurement of the kinetic SZ effect on many clusters using an 
optimal filtering technique would make a measurement of very large 
scale velocity flows possible~\cite{Haeh96,Agha97}.
\item The SZ effect is the strongest ``contamination'' source for the
 measurement
of the primary CMB anisotropies at high angular resolution and in the
millimetre spectral window, and therefore deserves careful studies (one's
noise is the other's signal), especially in the light of the
preparation to the Planck mission~\cite{COSA96}. 
\end{itemize}

In an effort to detect the SZ effect in clusters at high redshift, we
installed the Diabolo photometer at the focus of the IRAM 30\metre{}
millimetre radiotelescope (MRT). This photometer saw its first light
(Benoit \etal{} \cite{Beno}) at the Millimetre Infrared Testa Grigia
Observatory (MITO) in Italy on a 2.6\metre{} telescope.  The task of
detecting a signal which is a part in a million of the background is
very challenging but at a wavelength around 2\milli{}, the confusion
by other astrophysical sources (dust, point sources, CMB
anisotropies~\cite{Fran91,Fisc93}) is minimal. In addition, the high
angular resolution achieved with the 30\metre{} facility (about 30
arcseconds for the two Diabolo channels) reduces the beam dilution on
distant clusters.  Owing to major improvements in bolometer and
cooling technology, this task can now be achieved in a reasonable
integration time (a few hours). The observations and data reduction
method are described in section~\ref{se:obse}, and the results are
presented and discussed in section~\ref{se:resu}.

\section{Observations at the IRAM 30 m telescope}\label{se:obse}
The Diabolo experiment is a dual--channel photometer of which the
innovative cooling system, bolometers and readout electronics are
prototypes for space submillimetre astronomical applications (the ESA
Planck Surveyor mission~\cite{COSA96} and FIRST
cornerstone~\cite{Pill97}).  It is a cryostat with two bolometers
observing around 1.2 and 2.1\milli{}, cooled to 0.1 K by a
\nuc{3}{He}--\nuc{4}{He} compact dilution fridge. The two bands
matching the atmospheric windows are obtained with low pass filters
common to the two channels and free-standing bandpass meshes after the
light is selected by a dichroic beam splitter. The bolometer at
1.2\milli{} provides a constant monitoring of the so-called
atmospheric noise in a co--aligned and co--extensive beam with respect
to the 2.1\milli{} ``astrophysical'' bolometer channel. The
instrument, described in length by Beno\^{\i}t \etal{}~\cite{Beno},
has been modified as follows for the present observations:

\begin{itemize}
\item Only one
bandpass filter is used for the 2.1\milli{} channel, instead of two, in order
to increase the detection efficiency. We checked that the small spectral
leaks that appeared at high frequency have no influence on the SZ 
measurements. 
\item New readout electronics, now fully  digital, have been used.
Each bolometer is AC square--wave modulated in opposition in a Winston bridge
with a stable capacity. The out-of-equilibrium voltage is amplified by a cold
FET and warm amplifiers, AD converted, 
and then numerically demodulated after the
electrical transients have been blanked.
The digital signal is proportional to the total power
received by the bolometer up to an arbitrary offset constant. A
complete discussion of the readout electronics scheme can be found 
in~\cite{Gaer97}.
\item A NbSi thermometer has been installed on the dilution
fridge to monitor the $100\, {\rm mK}$ cold base plate temperature. 
Another resistance used as a heater now allows 
an active regulation of this base plate
temperature within about $30 \mu\rm K$. This is especially useful for
skydips (see Paragraph~\ref{se:skd}) and to avoid changes in the responsivity.
\item A warm polyethylene lens (90\% transmission) has been installed
in front of the cryostat to match the f--ratio of the telescope (about ten)
with that of the instrument (about five). 
\end{itemize}

The photometer has been installed at the Nasmyth focus of the
telescope for a test run from November 10th to November 14th 1995,
when the precipitable water vapor was too large (typically 5 to 9
\milli{}) for sensitive measurements.  The sensitivity and calibration
of the instrument could nevertheless be measured on bright sources.
Some 100 hours of observing time were allocated from December 1st to
4th from which the following results have been obtained. These
observations were complemented with a few more hours in December 1996.
                         

\subsection{Calibration}
\subsubsection{Alignment} 

The alignment of the cryostat with respect to the telescope axes was
achieved using a movable hot load situated between the entrance of the
cryostat and the secondary. The recording of the signal in total power
mode gives the beam direction and the appropriate corrections to be
done for the cryostat optical axis to be pointed at the center of the
secondary, which is crucial for straylight minimisation.

\subsubsection{Pointing}

\begin{figure}[t]
\centerline{\psfig{figure=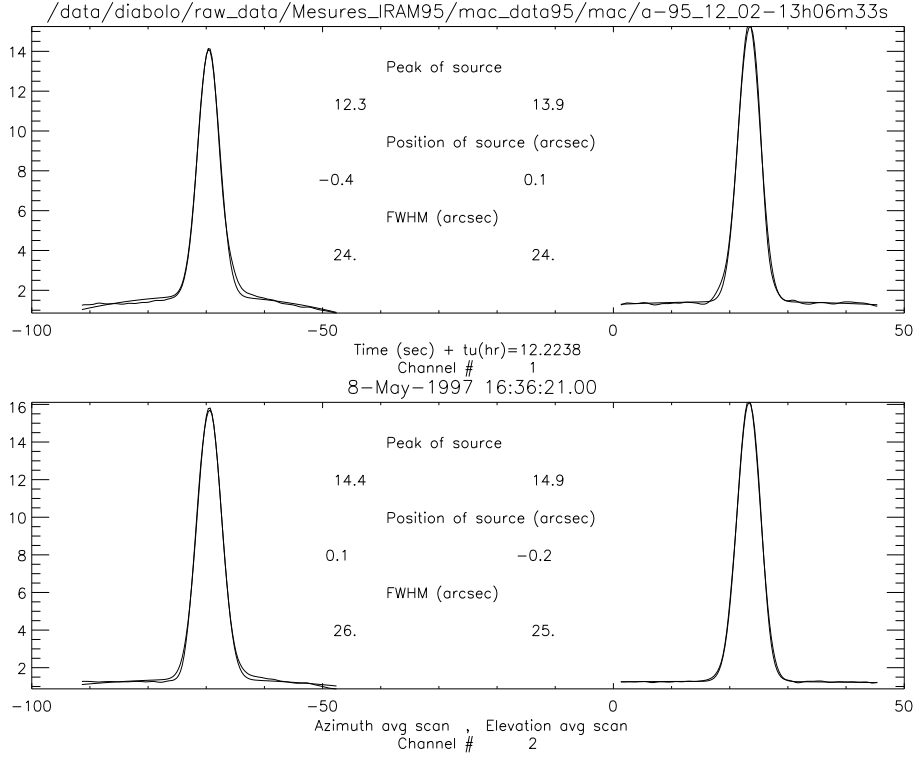,height=10.cm,width=12.cm,angle=0}}
\caption{The average demodulated signal from back and forth scans
across Mars. The top and bottom panels are for the 1.2\milli{} and
2.1\milli{} channels respectively. 
For both panels, the left plot corresponds to a constant elevation scan,
the right one to a constant azimuth scan. The X axis is in arcseconds,
the Y axis is in $\mu{\rm V}$. A Gaussian fit is superposed to the
data. Parameters of the fit are written on each panel. The precision on
the center and FWHM is about 2 and 3 arcseconds respectively.}
\label{fig:pointing}
\end{figure}

Pointing corrections were made every two hours, using data obtained by
scanning across a strong (several Jy) source (planet, quasar) situated
near the target. The signal was modulated by the wobbling secondary at
about 1 Hz. Fig.~\ref{fig:pointing} shows the demodulated signal as a
function of telescope direction along lines of constant elevation and
constant azimuth. A Gaussian fit is made to determine pointing
corrections if necessary.

\subsubsection{The beam pattern} 

\begin{figure}[t]
\epsfxsize=12.cm
\epsfysize=10.cm
\epsfbox{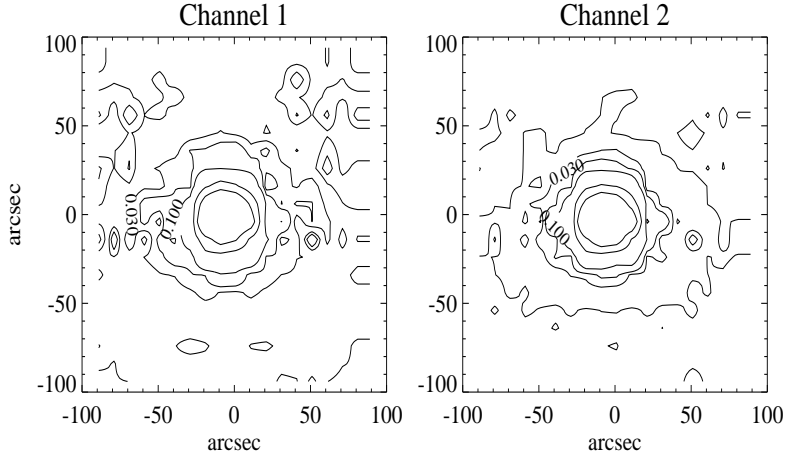}
\caption{Contour map of the two
beams observed by mapping Saturn. Contour levels are at 0, 1, 3, 5,
10, 30, and 50\% of peak value.}
\label{fig:beama} 
\end{figure}

\begin{figure}[t]
\centerline{\psfig{figure=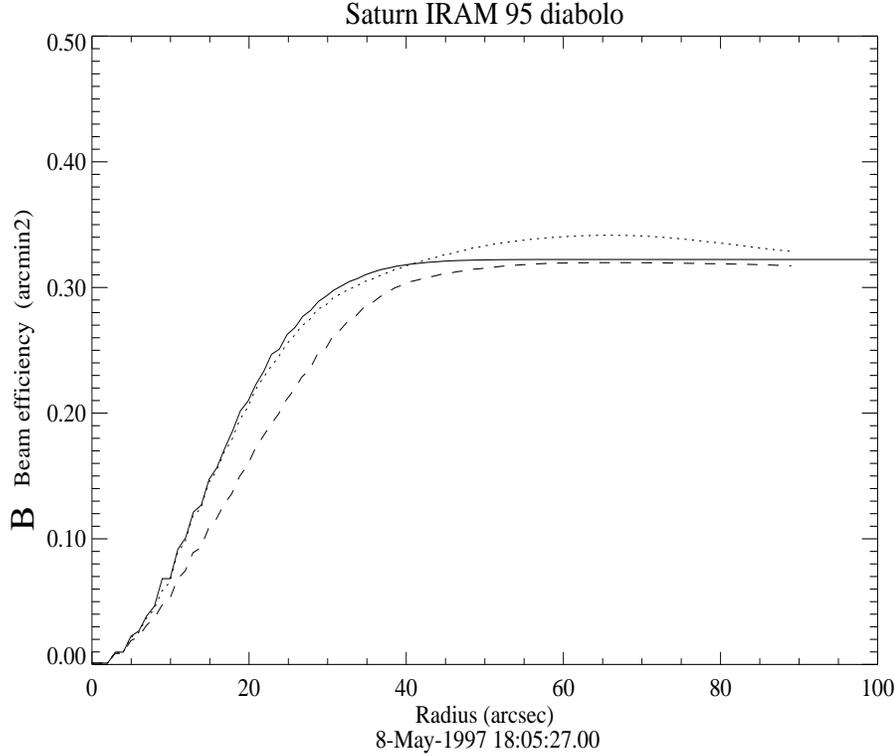,height=10.cm,width=12.cm,angle=90}}
\caption{ Integrated beam efficiency as observed with
Saturn. The function $B(\theta)$ defined by Eq.~\ref{eq:beam} is
plotted for the two wavelengths (dotted line for 1.2\milli{} and
dashed line for 2.1\milli{}) against the angular
radius $\theta$. The simulated integrated beam of a 2-dimensional
Gaussian of FWHM 34 arcsecond convolved with the Saturn 17 arcsecond
disk is shown as a plain line.}
\label{fig:beamb}   
\end{figure}

The beam pattern has been measured on Saturn in the November 95 test
run with a simple azimuth-elevation mapping technique.  It is shown in
Fig.~\ref{fig:beama} for the two wavelengths. The beam centers (as
defined by Gaussian one-dimensional fits) are within less than 2
seconds of arc from each other, confirming the accuracy of optical
positioning of the two bolometers with respect to the system optical
axis inside the cryostat.  Fig.~\ref{fig:beamb} shows the two
integrated beam profiles as defined by the function of the angular
radius $\theta$ starting from the center of the beam:
\begin{equation}
B(\theta)=\int_0^\theta {\rm d}\theta'\, \theta'\int_0^{2\pi} {\rm d}\phi
\frac{S(\theta',\phi)}{S(0,0)} \zu , 
            \label{eq:beam}
\end{equation}
where the measured signal $S$ is in cylindrical coordinates and where
an offset, estimated in the outskirts of the beam ($\theta > 45 \zu
arcsec$) has been taken out. $B$ has units of a solid angle and
represents the integrated beam efficiency which levels off at large
$\theta$. The beams for the two channels are much alike, except that
the longer wavelength channel one is slightly more extended because of
diffraction effects. Saturn is not point-like (17 arcsec diameter) and
slightly distorts the real beams.  The integrated beamwidth, calculated
from the integrated beam solid angle $\Omega_{\rm mb}$ as
$\theta=\sqrt{4\Omega_{\rm mb}/\pi}$ is larger than the
one--dimensional Gaussian FWHM (34 instead of 25 arcseconds), because
of near--sidelobe wings.


\subsubsection{Skydips}\label{se:skd}
Skydips must be performed in order to compare fluxes
measured at different elevations $\beta$. If the optical depth at the zenith
$\tau_0(\lambda)$ is known, all the measurements
 $F$ can be put on the same scale
``outside'' the atmosphere, yielding corrected measurements $F_c$. 
Assuming a plane-parallel geometry, this can be written as:
\begin{equation}
F_c(\lambda)=F(\lambda) \exp\left({\tau_0(\lambda)\over\sin\beta}\right)\zu .
            \label{eq:skf}
\end{equation}
Skydips were done in total power mode without any modulation, by having a
scan of the whole telescope at constant azimuth through 10 steps of
elevation with a constant cosecant increment.
The skydip technique has been pionneered by Chini \&
Kreysa~\cite{Chin86} at the IRAM 30\metre{} telescope. 
Here, we did not need a chopper for reference. 
The signal $S_i$ in a given channel and at
elevation $\beta_i$, for an average
atmospheric temperature $T_{atm}$, can be written as:

\begin{equation}
S_i= 
      C+ B_f
      T_{atm}\left(1-\exp\left(-{\tau_0\over\sin\beta_i}\right)\right)\zu .
\label{eq:skd}
\end{equation}

\begin{figure}[t]
\centerline{\psfig{figure=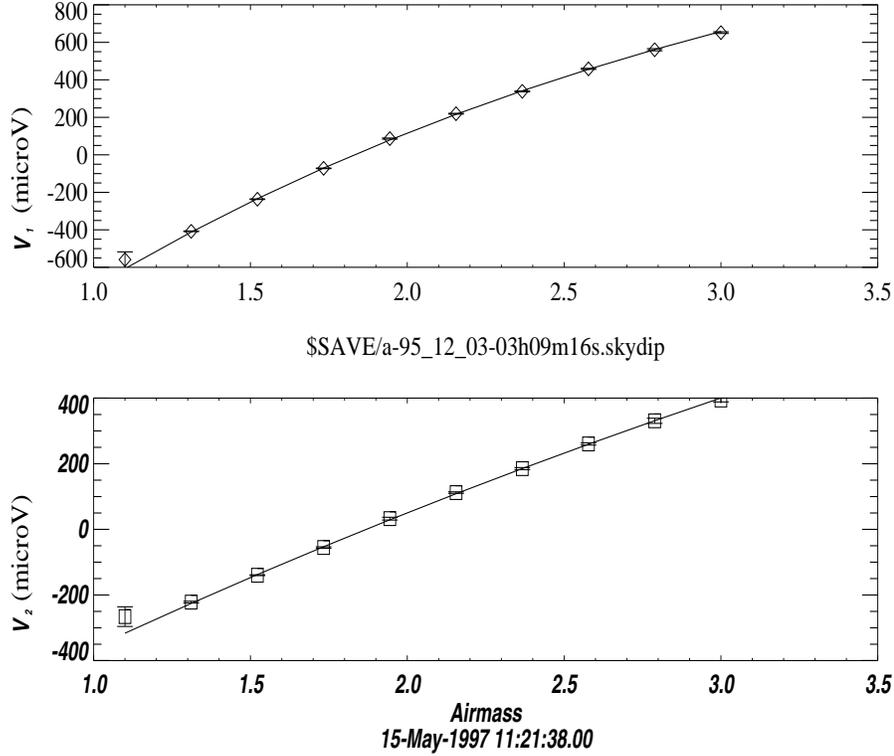,height=10.cm,width=12.cm,angle=90}}
\caption{ Total power skydip measurements for the 2 channels. The
average signal output in $\mu {\rm V}$ as a function of airmass is
fitted using the model of Eq.~\ref{eq:skd}. 
}
\label{fig:skd}   
\end{figure}

At each wavelength (1.2 and 2.1\milli{}), the constant $C$ represents
an arbitrary zero level. The forward beam efficiency $B_f={\rm d}
S/{\rm d} T$ is compared to the main beam efficiency $B_m$ (see below
Paragraph~\ref{se:sens}).  Before formula~\ref{eq:skd} can be applied,
one has to correct for the drifts of the $100\,{\rm mK}$ base plate
temperature $T_{bath}$, induced by the increasing heat load that
occurs with the skydip. The NbSi thermometer gives a sufficiently
sensitive measurement of $T_{bath}$. With a simple linear correlation
technique, the coefficients of which are established independently of
the skydip, the contribution ${\rm d} S/{\rm d} T_{bath}\times
T_{bath}$ can be subtracted.
Fig.~\ref{fig:skd} shows the non-linear fit of the data based on
formula~\ref{eq:skd}. 
The correction which
is applied to the data via Eq.~\ref{eq:skf} is deduced by interpolating
between the two observed skydip values of $\tau_0(1.2{\zu mm})$ closest in
time to the observation. It is found to be only of the order of 30\% or less
at 2.1\milli{} where the SZ effect is expected. 
During the December 95 observations, the zenith optical depths at
1.2\milli{} varied between 0.1 and 0.3, which corresponds to 2--4
\milli{} of precipitable water vapour. This definitely is an 
acceptable range of opacity values for SZ measurements.     

\subsubsection{Sensitivity}\label{se:sens}
The calibration is done with planets which partially fill the beam.
Mars (angular diameter 4.1 arcsec.), Saturn (16.7 arcsec.) and Jupiter
(30.5 arcsec.) have been used for the present observations assuming a
blackbody emission with temperatures of respectively 214, 150, and
170$\,{\rm K}$.  After correcting for atmospheric opacity effects
(Eq.~\ref{eq:skf}), and taking into account the beam dilution, the
responsivity of each bolometer $B_{mc}(\lambda)={\rm d} S/\d T$ is
deduced. It represents the response of the bolometer to
1~Rayleigh-Jeans Kelvin filling the main beam.  The noise level is
measured on blank fields. The instrument noise on the sky was found to
be above the bolometer noise (as measured in the laboratory) by a
factor of 3. The additional noise is likely related to an imperfect
isolation from vibrations in the Nasmyth cabin, which generates noise
of microphonic origin by optical modulation of straylight.

The final sensitivities are given in Table~\ref{ta:sens}, when the sky
noise is minimal and the zenith opacity is 0.1 at 1.2\milli{}. FWHM is
given by the point source profile Gaussian fit and $\Omega_{\rm mb}$
is the integrated beam solid angle up a 45 arcseconds radius.
Brightness sensitivities are for a filled beam, and flux sensitivities
are for a point source on axis.  These best performances are degraded
whenever the source is not at the zenith, the sky is less transparent
or the sky is more noisy.  The overall noise degradation can be by as
much as a factor of 3 at 1.2\milli{}, but rarely exceeds fifty percent
at 2.1\milli{}. In all cases, sky noise can be reduced by a
decorrelation technique (see Section~\ref{se:resu}). The corresponding
noise levels are given in parentheses in Table~\ref{ta:sens}.  For the
observation technique described in section~\ref{se:os}, the effective
sensitivity is worse than in Table~\ref{ta:sens} by a factor of 2.

\begin{table}
\caption{Best sensitivities obtained with Diabolo at the IRAM
30\metre{} telescope in 1995. Sensitivities in parentheses are for the
2.1 mm channel after spectral decorrelation of the atmospheric noise
(see text).}\label{ta:sens}
\begin{tabular}{lrrrrr}
\hrulefill&\hrulefill&\hrulefill&\hrulefill&\hrulefill&\hrulefill\\
Channel     & FWHM arcsec. & $\sqrt{\frac{4\Omega_{\rm mb}}{\pi}}$ arcsec.
 & mK.s$^{1/2}$ & MJy\,sr$^{-1}$.s$^{1/2}$ &
                 mJy.s$^{1/2}$ \\
\medskip\hrulefill&\hrulefill&\hrulefill&\hrulefill&\hrulefill&\hrulefill\\
\medskip
1.2\milli{} & $24\pm 3$ & $34 \pm 2$ & 25 & 50 & 900\\
2.1\milli{} & $27\pm 3$ & $34 \pm 2$ & 13(11) & 8(7) & 170(140)\\
\hrulefill&\hrulefill&\hrulefill&\hrulefill&\hrulefill&\hrulefill\\
\end{tabular}
\end{table}

The ratio between the corrected main beam efficiency $B_{mc}(\lambda)$
(obtained from mapping planets) and the forward beam efficiency
$B_f(\lambda)$ (obtained from skydips: Eq.~\ref{eq:skd}) is only 25\%
$\pm 5$ (50\% $\pm 5$) at 1.2 (resp. 2.1)\milli{}. These values are in
agreement with the telescope efficiencies measured by Garcia-Burillo
\etal{}~\cite{Garc93}. The far sidelobe pattern implied by these
results can be troublesome for the observation of weak sources. This
question is addressed in the discussion of Section~\ref{se:resu}.

\subsection{Observing strategy}\label{se:os}
Four types of modulation were simultaneously used in order to
limit the various low--frequency noises and monitor systematics.

\begin{enumerate}

\item The electronic AC modulation, referred to in the beginning of this
section, avoids using electronics at frequencies below $10\zu Hz$ (the
typical $1/f$ knee frequency). 
Here we modulate the bolometers at $36\zu Hz$ and the
readout electronics deliver one sample per bolometer at the rate of
$72\zu Hz$. 

\item The wobbling secondary provides the second modulation at the
  typical frequency of $1\zu Hz$ and with a beamthrow of 3 arc
  minutes.  This allows the slowly varying background emission (sky
  and telescope) to be subtracted from the comparison of the on-axis
  measurement with that from an offset position at the same elevation.
\item The whole telescope is nodded in azimuth every 20 seconds with
  an amplitude of 3 arcminutes in an ABBA cycle which is repeated 4
  times to form one scan. This permits to compensate for any imbalance
  between the two beams provided by the wobbling.  Each scan obtained
  in this way lasts about 2 minutes (repointing overheads included).
\item Each scan above is done consecutively on a reference field
  offset from the target by a lag of a few minutes of time in RA ($R$
  at coordinates $(\alpha - {\rm lag}, \delta)$), on target twice ($T$
  and $T'$ at coordinates $(\alpha, \delta)$), and on a second
  reference field offset by the same number of minutes of time in RA
  in the other direction ($R'$ at coordinates $(\alpha + {\rm lag},
  \delta)$). With this method, the reference fields are followed in
  the same way as the target in local coordinates.  This ensures that
  sidelobe effects (ground pickup), if any, are subtracted.  This
  technique has been used by Herbig \etal{}~\cite{Herb95} for
  single-dish measurements of very weak sources with proper baseline
  subtraction.

\end{enumerate}

\subsection{Reduction procedure}\label{ss:red}

The data reduction proceeds as follows.

\begin{figure}[t]
\epsfxsize=12.cm
\epsfysize=10.cm
\epsfbox{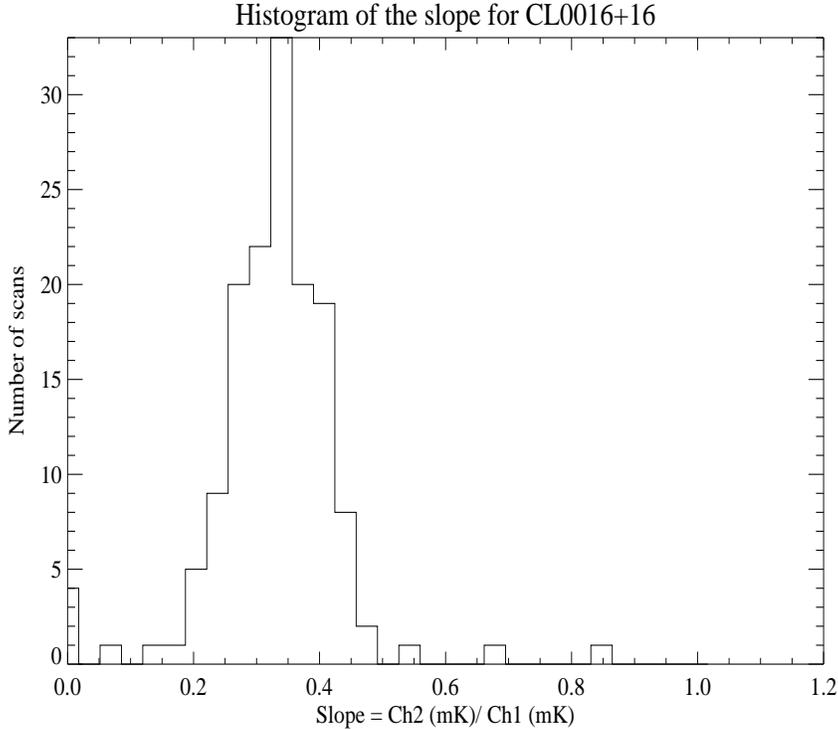}
\caption{Histogram of the values $r$, the slope between channel
2 against channel 1 in units of a temperature brightness ratio. A value of
0.3 is expected from the fluctuations of water vapour and the photometric
model of the instrument. Sky noise seems indeed to be the source of extra
noise seen in the bolometers.} \label{fig:atmos}    
\end{figure}

\begin{enumerate}
  
\item Cosmic ray hits are removed by interpolation from the data flow
  by a running median algorithm. Typical time constants are 10
  milliseconds and the glitch rate is less than one hit per bolometer
  every 10 seconds, so that few samples are affected. The particles
  which deposit their energy into the bolometer are thought to mainly
  be muons, more abundant at the telescope site than in the
  laboratory.
  
\item The data are then synchronously demodulated with the help of the
  wobbler position (which is recorded along with the bolometer
  signals). The mean and dispersion values are computed for each
  position of the nodding cycle $ABBA$. Typical offsets (the imbalance
  between the positive and negative wobbler positions) are of the
  order of $0.2\zu K$.
  
\item A complete scan is reduced by averaging the differences of
  values between the two nodding positions:
  $v=\sum_1^N(v_A-v_B)/(2N)$. The noise on the final value is obtained
  from the dispersion of the individual differences. The sensitivity
  quoted in Table~\ref{ta:sens} corresponds to the best noise figure
  obtained after a scan and corrected for the square root of the scan
  integration time.
  
\item The third channel, which is a thermometer measuring the base
  plate temperature $v_3=T_{bath}$, is treated the same way as the two
  others in order to check for a possible systematic effect or
  additionnal noise possibly induced by drifts of the thermal bath
  temperature.  None have been found.
\item A linear combination of the first two bolometers, $v_4=v_2- r
  v_1$ is calculated.  The ratio $r$ is chosen so as to minimise the
  noise of $v_4$. It can be shown that $r$ can be deduced from a
  simple linear correlation between $v_1$ and $v_2$ even if both
  measurements are noisy, and $r$ is always smaller than the color of
  the sky emission.  This procedure is intended to specifically work
  at removing sky noise from the second channel when little or no
  signal is expected from the first one (in particular, in case of the
  SZ effect). An histogram of the values of $r$ during the December
  observations is shown in Fig.~\ref{fig:atmos}. The correlation
  coefficient $C$ tells us by how much we can reduce the initial noise
  of $v_2$ to that of $v_4$: $\sigma_4=\sigma_2\sqrt{1-C}$. The
  typical correlation coefficient $C$ of 0.4 leads to a small
  improvement in the signal to noise ratio of weak sources. On the
  other hand the statistical distribution which is obtained with the
  corrected $v_4$ is much closer to Gaussian than that of the data for
  the original channels, $v_1$ and $v_2$ (see~\ref{ss:res}).
  
\item For each channel, an elementary block of data, made of 4 scans
  ($R$, $T$, $T'$, $R'$), is reduced by computing an average signal
  $s= (v_T+v_{T'}-v_R-v_{R'})/2$ and a difference signal $d=
  (v_{T'}-v_T-v_R+v_{R'})$ with associated errors.

\end{enumerate}

Four rich clusters of galaxies (A665, A2163, A2218 and CL0016+16) have
been selected for observations due to their small angular core radius
(less than 2 arcminutes), adapted to a large millimetre antenna.  In
addition to the 1995 data, we gathered a few more hours of observation
towards two of these clusters (A2163 and CL0016+16) in 1996.  The
observation and data reduction schemes were very different, in an
attempt to measure SZ profiles.

\section{Tentative detections of the SZ effect}\label{se:resu}

\subsection{Results}\label{ss:res}

\begin{figure}[p]
\epsfxsize=12.cm
\epsfysize=10.cm
\epsfbox{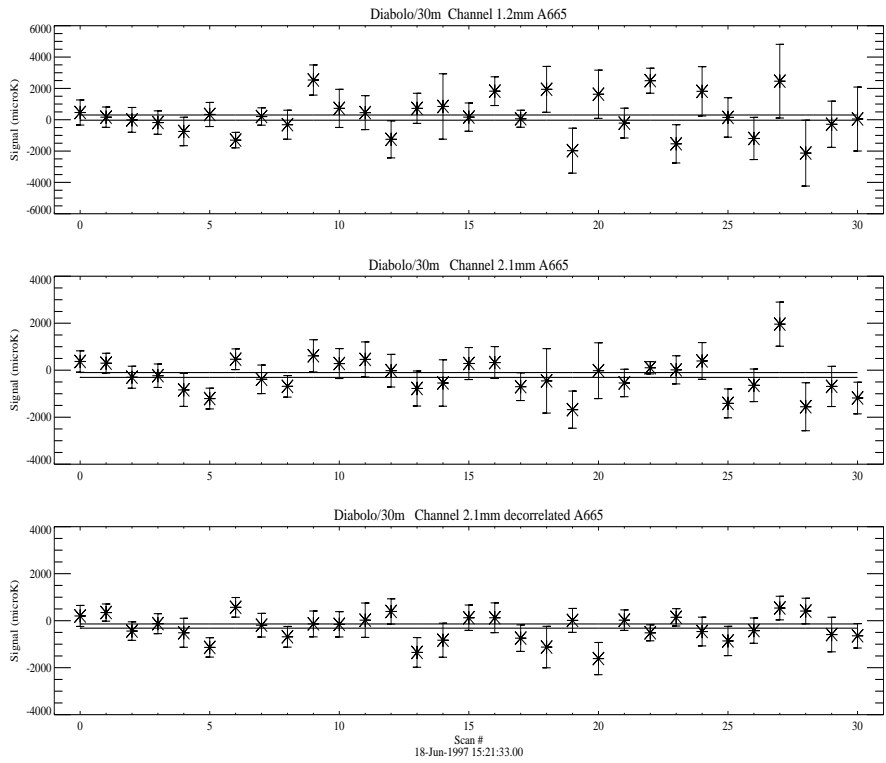}
\caption{Antenna temperature observed for cluster A665.  
The two upper plots show the two independent 
diabolo channel measurements and the lower plot is the the second
channel corrected for atmospheric noise (see text). The 2 lines show
$\pm 1 \sigma$ from the final optimally averaged value.} 
\label{fig:res665}     
\end{figure}

\begin{figure}[p]
\epsfxsize=12.cm
\epsfysize=10.cm
\epsfbox{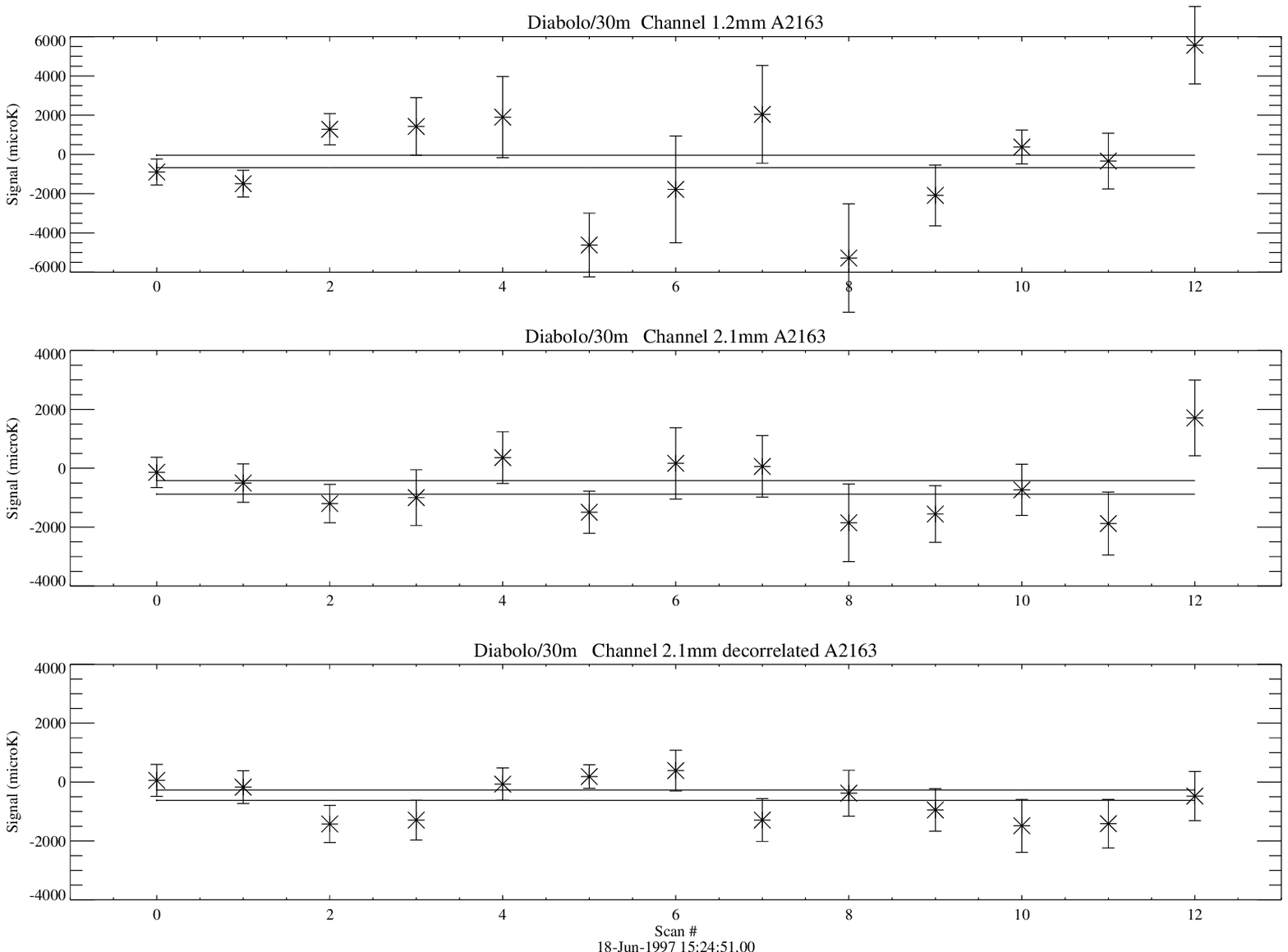}
\caption{Antenna temperature for the cluster A2163.}
\label{fig:res2163}    
\end{figure}

\begin{figure}[p]
\epsfxsize=12.cm
\epsfysize=10.cm
\epsfbox{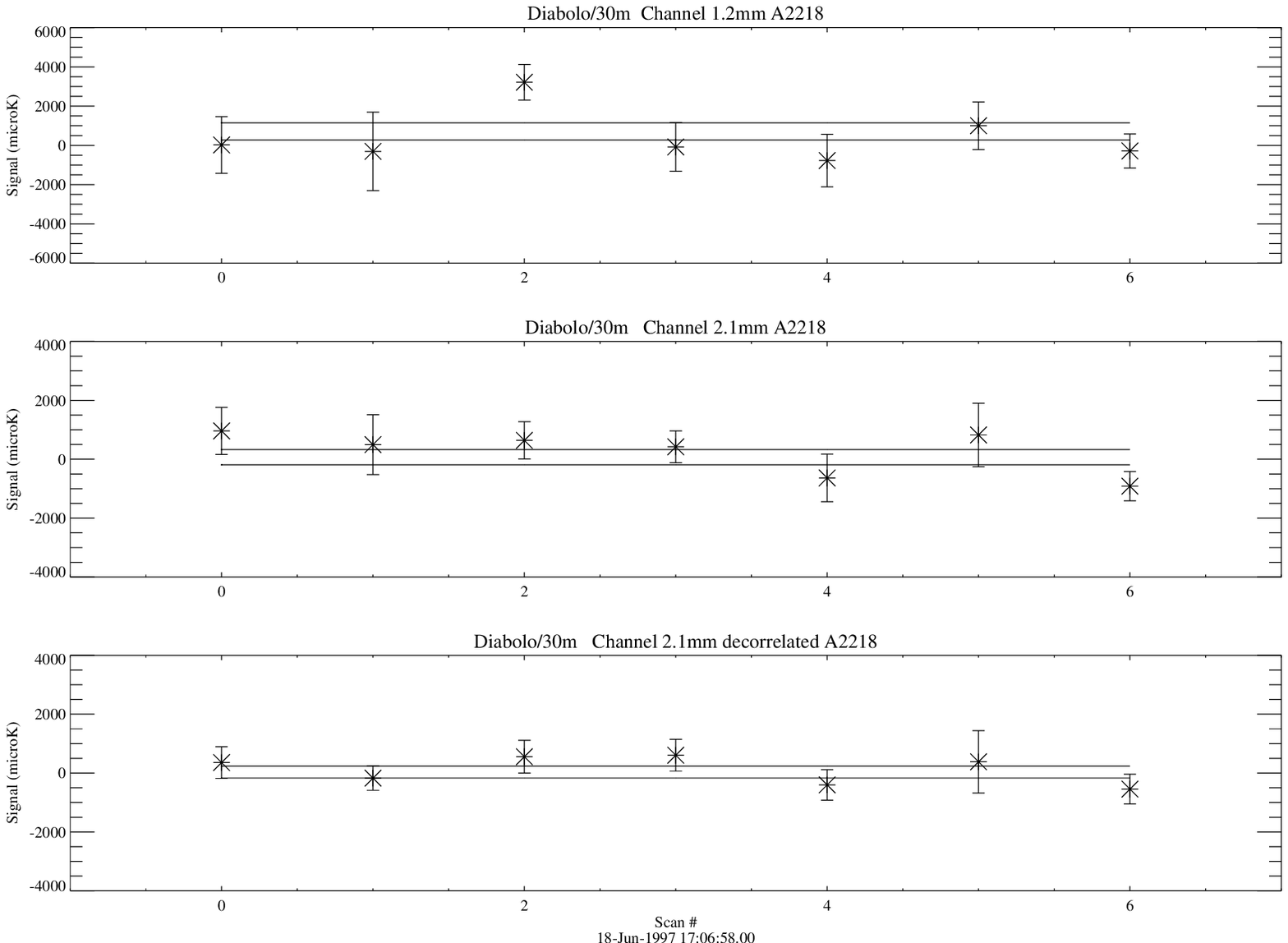}
\caption{Antenna temperature for the cluster A2218.}
\label{fig:res2218}    
\end{figure}

\begin{figure}[p]
\epsfxsize=12.cm
\epsfysize=10.cm
\epsfbox{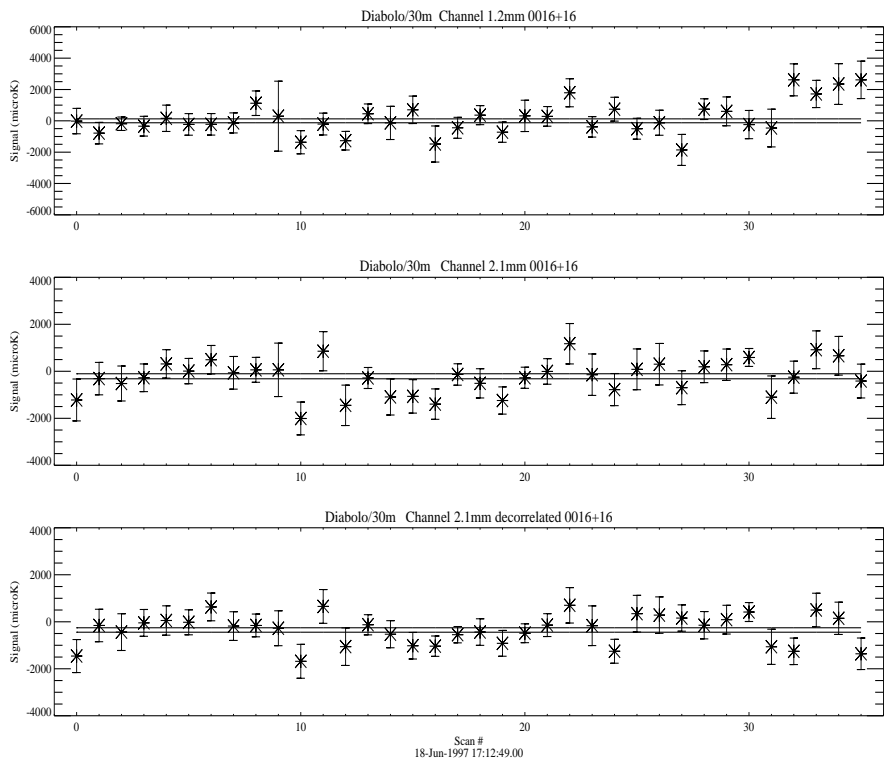}
\caption{Antenna temperature for the cluster CL0016+16.} 
\label{fig:res0016}     
\end{figure}

The parameters of the observations towards the four clusters are given
in Table~\ref{ta:log}, and the full results are summarised in
Table~\ref{ta:res1} and~\ref{ta:res2} (antenna Rayleigh-Jeans
equivalent temperature in \microK, corrected for atmospheric
absorption).  Rayleigh-Jeans temperature differences (the corrected
signal $s$ see~\ref{ss:red}) for all cycles of measurements are
plotted in Fig.~\ref{fig:res665} to Fig.~\ref{fig:res0016}.

For the 1995 data, the final result for each cluster has been obtained
by averaging the measurements obtained from each cycle of four scans,
weighted proportionally to the inverse square of the noise for the
individual sets. For each measurement, we compare the internal error
obtained with this optimal averaging by using the internal noise
value, and the external error obtained from the dispersion between the
scan values.  The square of the ratio between the two is the reduced
$\chi^2$.  The values listed in Tables~\ref{ta:res1} and \ref{ta:res2}
show the internal consistency of the measurement and its estimated
noise, except for the first channel where the $\chi^2$ value is
systematically larger than unity. This discrepancy can be explained by
the statistics of the atmospheric noise, which is not Gaussian and
affects more the 1.2\milli{} channel than the 2.1\milli{} one.  To
first order, it should not affect much the decorrelated channel, as
observed.

For the 1996 data, which use a different scanning technique, results
have been obtained from the difference between the average of the
signal from the scans on 30 arcseconds centered on the target source
and the average value of the signal at more than 40 arcseconds of the
target.

A significant negative signal is detected in the 2.1\milli{}
decorrelated channel for the three clusters A665, A2163 and CL0016+16.
This detection is particularly significant for the latter cluster.

If we interpret those measurements as due to the Sunyaev-Zel'dovich
effect, one can convert the obtained value from antenna temperature to
the $y$ parameter (see~\cite{Suny70,Suny80}), neglecting the spectral
dependency on the cluster gas temperature (\cite{Reph95,Giar95}).  The
final results (1995, 1996, and combination of the two years) are given
in Table~\ref{ta:sz}.

The correction $\eta$ for the 1995 beam dilution is calculated by convolving 
the measured beam profile obtained on Saturn  (modulated at 3 arcmin)
with a theoretical SZ profile using core parameters from X-ray 
measurements. 

We do not detect any significant signal in the blank field positions
($v_R+v_{R'}$): in contrast to cm radio observations, no systematic
signal is seen in the blank field measurements. Indeed, we find that the
average signal obtained by keeping only the on-source component
($v_T+v_{T'}$) is about $\sqrt{2}$ more significant than that shown in
Tables~\ref{ta:res1} and \ref{ta:res2}. This gives us confidence in
the final results for the $y$ parameters of Table~\ref{ta:sz}.
Moreover, for A2163 and CL0016+16, the 1995 and 1996 results are
compatible with each other.

An additional outcome of the observations are temperature differences
from blank field measurements ($d$: see~\ref{ss:red}). No signal is
detected in any of the 4 differences (around the 4 observed clusters)
at the $\Delta T/T= 2\times 10^{-4}$ level ($1\sigma$). If this result
were improved by repeated measurements on a larger number of clusters,
it could yield interesting constraints on the level of CMB
anisotropies at small angular scales (30 arcseconds to few arcminutes)
in a wavelength range where the smallest contamination from radio and
infrared galaxies~\cite{Fran91} is expected.

All clusters show no signal at 1.2\milli{} within the observational
errors. In principle, both the thermal and kinetic SZ effect could
contribute to this channel, but the upper limit that we can put on
cluster radial velocities is not stringent enough to be relevant. No
galactic dust emission is detected either. 

\begin{table}
\caption{Final calibrated results}
\label{ta:sz}
\begin{tabular}{|l|cc|c|cc|}

\hline
        & & 1995  &  1996  & all & \\
Cluster  & $\eta $    & $y_0$ $\, /10^{-4}$ &  $y_0$ $\, /10^{-4}$ 
& $y_0$ $\, /10^{-4}$ & S/N  \\
\hline
A665    & 0.498 &  $2.92\pm 1.15$ &                  & $2.92\pm 1.15$ 
& 2.5\\
A2163   & 0.548 &  $4.99\pm 1.97$ & $4.60\pm 2.00$ & $4.80\pm 1.40$
& 3.4\\
A2218   & 0.607 &  $-0.37\pm 2.21$ &               & $-0.37\pm 2.21$
& 0.2\\
CL0016+16 & 0.668 &  $3.30\pm 0.90$ & $2.90\pm 1.60$ & $3.20\pm 0.78$
& 4.1\\ 
\hline
\end{tabular}
\end{table}

\begin{table}
\caption{Physical parameters of the observed clusters taken from
\cite{Birk91b,Elba95,Birk94,Neum97,Hugh95}. The total gas 
mass is computed from the
present measurements. Uncertainties are statistical only.}
\label{ta:amas}
\begin{tabular}{|l|cccc|cc|}

\hline

Cluster  & $z $    & $T_e\, ({\rm keV})$ & $\theta_c$ arcmin &  $\beta$ & 
$Y \; (10^{-4}\,{\rm arcmin}^2)$ & $M_g / 10^{14}M_\odot$ \\
\hline
A665    & 0.182 & 8.2   & 1.60  & 0.66  & 439   & $20.2 \pm 8.0$ \\
A2163   & 0.201 & 14.6  & 1.20  & 0.62  & 491   & $14.6 \pm 4.2$ \\
A2218   & 0.171 & 6.72  & 1.00  & 0.65  & $< 408$ & $< 20.9 \,(3\sigma)$\\
CL0016+16 & 0.541 & 8.22        & 0.64  & 0.68  & 70    & $11.1 \pm 2.7$ \\
\hline
\end{tabular}
\end{table}

\subsection{Interpretation}

The mass of hot gas can be directly deduced from these
observations by using:

\begin{eqnarray}
M_g= 8.2\times 10^{14}M_\odot
  \left(h\over 0.5 \right)^{-2} 
  \left({Y \over {10^{-4} {\zu arcmin^2}}}\right) \nonumber\\
        \left({k T_e} \over {10 {\zu keV}} \right)^{-1}
          {(\sqrt{1+z} -1)^2 \over {(1+z)^3}} \zu , 
            \label{eq:gasm}
\end{eqnarray}
a formula derived by De Luca \etal{}~\cite{Delu95}. Here we have assumed
 $\Omega_0=1$ and
$h= H_0 / (100 \zu km/s/Mpc)$, and the measurement $y_0$ has been converted
into $Y = \int y d\Omega = y_0 
\Omega_{\rm eff}$. The effective solid angle $\Omega_{\rm eff}$
is calculated with
\begin{eqnarray}
{\Omega_{\rm eff}\over{\theta_c^2}}= 
f_{\rm geom}= 2\pi\int x {\rm d}x(1+x^2)^{(1-3\beta)\over 2}\, , 
\end{eqnarray}
and $x=\theta/\theta_c$, assuming a King profile with an angular core
radius of $\theta_c$. 
The resulting masses are given in Table~\ref{ta:amas}. Parameters for
the clusters, $\theta_c$, $\beta$, and $T_e$, have been taken from recent
ROSAT X-ray measurements. These estimated 
masses do not depend on the absolute X-ray fluxes. 

Our result for A2163, $y_0 = 4.8 \pm 1.4 \times 10^{-4}$ is in
agreement with the determination by Wilbanks \etal{}~\cite{Wilb94} of
$y_0 = (3.78^{+0.74}_{-0.65}) \times 10^{-4}$ and that of Holzapfel
\etal{}~\cite{Holz97} of $y_0 = (3.73^{+0.47}_{-0.61}) \times
10^{-4}$, both obtained at the same wavelength ($2.1\milli{}$) as the
present measurements with the 1.4' beam (2' throw) of the SuZie
experiment. It is also in agreement with the submillimeter detection
by the SPM photometer onboard the PRONAOS balloon (with a 3.7' beam
and 6' beamthrow).  A detailed discussion of the combined bolometer
results for A2163 is given by Lamarre \etal{}~\cite{Lama98}.  The gas
mass we deduce is $14.6 \pm 4.2 \times 10^{14} M_\odot$, very close to
the X-ray determined gas mass~\cite{Elba95} $14.3 \pm 0.5 \times
10^{14} M_\odot$.

Our most significant detection (at the 4 $\sigma$ level) concerns the
distant cluster CL0016+16 at a redshift of 0.541.  This cluster is the
highest redshift object detected with the SZ effect in the millimetric
domain.  Our result of $y_0 = 3.20 \pm 0.78 \times 10^{-4}$ is larger
than but compatible with the central parameter $y_0= 2.18\times
10^{-4} (h/0.5)^{-1/2}$ predicted by Birkinshaw~\cite{Birk98} using
ROSAT X-ray data (within $1.3\sigma$).  It is in agreement with the SZ
radio determination of Hughes and Birkinshaw~\cite{Hugh98} with a
larger beam (1.8' with a 7' beam throw) of $y_0 = 2.20 \pm 0.37 \times
10^{-4}$ (see also~\cite{Birk91a}), and more marginally with the SZ
map of the interferometer experiment of Carlstrom
\etal{}~\cite{Carl96} of $y_0 = 1.31 \pm 0.12 \times 10^{-4}$, which
spans 1 to 10' angular scales.  Our gas mass estimate of $M_g = 11.1
\pm 2.7 \times 10^{14} M_\odot$ is twice as large as the X-ray gas
mass deduced by Neumann and B\"ohringer~\cite{Neum97} but still within
errors.

For A665, the observations were centered on the IPC X-ray center as
given by Birkinshaw, Hughes \& Arnaud~\cite{Birk91b}, which is offset
by 2' from the nominal Abell center. Although less significant, the
measured central brightness decrement $y_0 = 2.92\pm 1.15 \times
10^{-4}$ is in agreement with the more accurate value $1.69\pm 0.15
\times 10^{-4}$ determined by Birkinshaw \etal{}~\cite{Birk91b},
albeit in the radio domain.

The integration time was clearly insufficient for A2218 to reach a
significant noise level for that cluster. The upper limit that we get
is compatible with the radio measurements that were previously
reported~\cite{Birk91a,Jone93}.

\subsection{Perspective}
We have reported here the highest angular resolution (30'')
observations of the SZ effect on at least 2 clusters. These
observations could be achieved thanks to the large millimetre Pico
Veleta antenna and a total on source integration time of fifty hours.
It is clear that SZ profiles or even maps of rich clusters can be
measured with the Diabolo instrument, with sufficient winter
integration time, when improvements in the overall efficiency are made
(these are currently underway). These observations are complementary
to X-ray measurements in the sense that they directly sample the gas
pressure with similar angular resolution (the future XMM and AXAF will
have few arcsecond resolutions). High resolution SZ observations in
the millimeter atmospheric windows will also grow in importance after
the unbiased survey of SZ clusters from the Planck Surveyor satellite.
For resolved clusters the amplitude of the SZ distortion is
independent of distance, and thus high-redshift clusters are adequate
targets for millimetre observations of the SZ effect, whereas X-ray
measurements of gas masses are more difficult.

\ack We wish to thank the IRAM staff especially for their help during
the setup of the instrument, Bernard Fouilleux for his help during the
observations, and Bernard Lazareff for his support of the mission. We
thank the whole Diabolo team for the continuous improvements brought
to the instrument, with a special attention to Jean-Pierre Crussaire,
Gerard Dambier, Jacques Leblanc, Bernadette Leriche, and Marco De
Petris along with the Testa Grigia MITO team for a previous test of
the instrument.  INSU, IAS, CESR, CRTBT, and the GdR Cosmologie
contributed financially to this instrument.

\begin{center}
\begin{table}
\caption{Observation Logbook Summary}\label{ta:log}
\begin{tabular}{lllrr}
\hrulefill\hrulefill&\hrulefill&\hrulefill&\hrulefill&\hrulefill\\
Cluster       &  RA (1950)   & Dec (1950) & Offset (arcsec)  & Int. time 
95 (96) \\
\medskip\hrulefill&\hrulefill&\hrulefill&\hrulefill&\hrulefill \\
A665 Source   &   08.26.25.0 & +66.01.21. &              0. & 13.1 hours\\
A665 Ref1     &   08.25.29.2 & +66.01.21. &           -340. &  \\
A665 Ref2     &   08.27.20.8 & +66.01.21. &           +340. &  \\
\medskip\hrulefill&\hrulefill&\hrulefill&\hrulefill&\hrulefill \\
A2163 Source  &   16.13.05.8 & -06.01.29. &              0. & 5.7 (3.2) \\ 
A2163 Ref1    &   16.13.05.8 & -06.01.29. &           -340. &  \\ 
A2163 Ref2    &   16.13.05.8 & -06.01.29. &           +340. &  \\ 
\medskip\hrulefill&\hrulefill&\hrulefill&\hrulefill&\hrulefill \\
A2218 Source  &   16.35.35.0 & +66.18.50. &              0. & 3.1 \\ 
A2218 Ref1    &   16.29.50.8 & +66.18.50. &          -2074. &  \\ 
A2218 Ref2    &   16.41.19.2 & +66.18.50. &          +2074. &  \\ 
\medskip\hrulefill&\hrulefill&\hrulefill&\hrulefill&\hrulefill \\
CL0016+16 Source&   00.15.58.3 & +16.09.37. &              0. & 15.7 (9.8)\\ 
CL0016+16 Ref1  &   00.15.34.7 & +16.09.37. &           -340. &  \\ 
CL0016+16 Ref2  &   00.16.21.9 & +16.09.37. &           +340. &  \\ 
CL0016+16 Ref1p &   00.10.18.2 & +16.09.37. &          -4900. &  \\ 
CL0016+16 Ref2p &   00.21.38.4 & +16.09.37. &          +4900. &  \\ 
\medskip\hrulefill&\hrulefill&\hrulefill&\hrulefill&\hrulefill \\
\end{tabular}
\end{table}

\end{center}

\begin{table}[p]
\begin{center}
\caption{Observation results}\label{ta:res1}
\begin{tabular}{llrrr}
\hrulefill\hrulefill&\hrulefill&\hrulefill&\hrulefill&\hrulefill\\
Channel     & $\lambda (\milli{})$ & Signal (\microK) (noise)
  & $\chi^2$ (df)  & Prob \%  \\
\medskip\hrulefill&\hrulefill&\hrulefill&\hrulefill&\hrulefill\\
A665 &&&&\\
\medskip
source - ref (microK RJ): 
  \hrulefill&\hrulefill&\hrulefill&\hrulefill&\hrulefill\\
1     & 1.2\milli{}     &     193.   (    235.    277.) & 41.6 (30) &  7.7 \\
2     & 2.1\milli{}     &    -224.   (    115.    129.) & 37.7 (30) & 15.9 \\
4     & 2.1\milli{} corr&    -253.   (    100.    106.) & 33.7 (30) & 29.3 \\
ref2 - ref1 (microK RJ): 
  \hrulefill&\hrulefill&\hrulefill&\hrulefill&\hrulefill\\
1     & 1.2\milli{}     &    -439.   (    470.    685.) & 63.8 (30) &  0.0 \\
2     & 2.1\milli{}     &     -80.   (    231.    298.) & 50.0 (30) &  1.2 \\
4     & 2.1\milli{} corr&     303.   (    200.    265.) & 52.8 (30) &  0.6 \\

\hrulefill&\hrulefill&\hrulefill&\hrulefill&\hrulefill\\
A2163 &&&&\\
\medskip
source - ref (microK RJ): 
  \hrulefill&\hrulefill&\hrulefill&\hrulefill&\hrulefill\\
1     & 1.2\milli{}     &    -451.   (    398.    656.) & 32.6 (12) &  0.1 \\
2     & 2.1\milli{}     &    -695.   (    247.    246.) & 11.9 (12) & 45.4 \\
4     & 2.1\milli{} corr&    -476.   (    188.    203.) & 14.1 (12) & 29.6 \\
ref2 - ref1 (microK RJ): 
  \hrulefill&\hrulefill&\hrulefill&\hrulefill&\hrulefill\\
1     & 1.2\milli{}     &   -1388.   (    797.   1282.) & 31.1 (12) &  0.2 \\
2     & 2.1\milli{}     &     961.   (    494.    657.) & 21.2 (12) &  4.7 \\
4     & 2.1\milli{} corr&     636.   (    376.    405.) & 13.9 (12) & 30.6 \\
 
\hrulefill&\hrulefill&\hrulefill&\hrulefill&\hrulefill\\
\end{tabular}
\end{center}
\end{table}

\begin{table}[p]
\begin{center}
\caption{Observation results: following} \label{ta:res2}
\begin{tabular}{llrrr}
\hrulefill\hrulefill&\hrulefill&\hrulefill&\hrulefill&\hrulefill\\
Channel     & $\lambda (\milli{})$ & Signal (\microK) (noise)
& $\chi^2$ (df)  & Prob
\%  \\
\hrulefill&\hrulefill&\hrulefill&\hrulefill&\hrulefill\\
A2218 &&&&\\
\medskip
source - ref (microK RJ): 
  \hrulefill&\hrulefill&\hrulefill&\hrulefill&\hrulefill\\
1     & 1.2\milli{}     &    1151.   (    712.    971.) & 11.2 ( 6) &  8.4 \\
2     & 2.1\milli{}     &      80.   (    299.    342.) &  7.9 ( 6) & 24.8 \\
4     & 2.1\milli{} corr&      39.   (    234.    207.) &  4.7 ( 6) & 58.1 \\
ref2 - ref1 (microK RJ): 
  \hrulefill&\hrulefill&\hrulefill&\hrulefill&\hrulefill\\
1     & 1.2\milli{}     &    3504.   (   1423.   2485.) & 18.3 ( 6) &  0.6 \\
2     & 2.1\milli{}     &       2.   (    598.    480.) &  3.9 ( 6) & 69.5 \\
4     & 2.1\milli{} corr&    -429.   (    467.    544.) &  8.2 ( 6) & 22.7 \\
  
\hrulefill&\hrulefill&\hrulefill&\hrulefill&\hrulefill\\
CL0016+16 &&&&\\
\medskip
source - ref (microK RJ): 
  \hrulefill&\hrulefill&\hrulefill&\hrulefill&\hrulefill\\
1     & 1.2\milli{}     &       1.   (    172.    200.) & 47.2 (35) &  8.1 \\
2     & 2.1\milli{}     &    -233.   (    118.    122.) & 37.8 (35) & 34.2 \\
4     & 2.1\milli{} corr&    -384.   (    104.    109.) & 38.2 (35) & 32.5 \\
ref2 - ref1 (microK RJ): 
  \hrulefill&\hrulefill&\hrulefill&\hrulefill&\hrulefill\\
1     & 1.2\milli{}     &     243.   (    344.    464.) & 63.6 (35) &  0.2 \\
2     & 2.1\milli{}     &     176.   (    235.    240.) & 36.4 (35) & 40.5 \\
4     & 2.1\milli{} corr&     -51.   (    209.    210.) & 35.5 (35) & 44.7 \\
   
\hrulefill&\hrulefill&\hrulefill&\hrulefill&\hrulefill\\
\end{tabular}
\end{center}
\end{table}

\end{document}